\documentstyle[eqsecnum,epsfig, aps,prd]{revtex}

\begin{document}
\draft
\preprint{hep-ph/0012217} 
\title{ Rare decays $B\to X_s\tau^+\tau^-$ and $B_s\to 
        \tau^+\tau^-\gamma$ in technicolor with scalars } 
         
\author{Zhaohua Xiong $^{a,b,c}$ and Jin Min Yang $^b$} 

\address{$^a$ CCAST (World  Laboratory), P.O.Box 8730, Beijing 100080, China}
\address{$^b$ Institute of Theoretical Physics, Academia Sinica, 
           Beijing 100080, China} 
\address{$^c$ Institute of High Energy Physics, Academia Sinica,
         Beijing 100039, China}
\date{\today}
\maketitle

\begin{abstract}
We examine the rare decays $B\to X_s\tau^+\tau^-$ and 
$B_s\to\tau^+\tau^-\gamma$ in the framework of technicolor with scalars. 
The contributions from both the neutral and charged scalars predicted in
this model are evaluated. We find that the branching ratios could be 
enhanced over the standard model predictions by a couple of orders of 
magnitude in some part of parameter space. The  forward-backward asymmetry
and the distributions of differential branching ratios are also found to 
differ significantly from the standard model results.
Such large new physics effects might be observable in the new generation 
of B experiments. 
\end{abstract} 

\bigskip  

\pacs{12.60.NZ,\ 13.25.Hw}

{\bf Key words}: Technicolor with scalars; Decay $B\to {X}_s\tau^+\tau^-$;
Decay $B_s\to\tau^+\tau^-\gamma$; Forward-backward asymmetry

\section{Introduction}

One intriguing puzzle in particle physics is the regular pattern of three 
lepton and quark families. The existence of families gives rise to many 
parameters of the standard model (SM).  Flavor-changing neutral-currents 
(FCNC) induced B-meson rare decays provide an ideal opportunity for 
extracting information about the fundamental parameters of the SM, such as 
the Cabibbo-Kobayashi-Maskawa (CKM) matrix elements, and for testing the SM
predictions at loop level and probing possible new physics. 
The experimental discovery of the inclusive and exclusive rare decays 
$b\to X_s\gamma$ and $B\to K\gamma$ \cite{CLEO9395} stimulated the study of 
radiative rare B-meson decays with a new momentum.

The inclusive decays $B\to X_s\ell^+\ell^-\ (\ell=e, \mu)$ have been well 
studied in the frameworks of  minimal supersymmetric model\cite{Grossman97}, 
the two Higgs doublet model(2HDM) \cite{Grinstein89,Dai97,Hewett96} and the 
technicolor models \cite{Weinberg79,Dimopoulos79}. It was shown that the 
matrix elements are strongly suppressed by a factor $m_\ell/m_W$ and the 
contributions from exchanging neural scalars can be safely neglected.
However, the situation is different in the case of $\ell=\tau$.  
The branching ratio $Br(B_s\to\tau^+\tau^-)\simeq 8\times 10^{-7}$
\cite{Buchalla93} in the SM is large enough to be observable   
in future B-factories. The contributions from  neutral scalars exchange to 
$B\to X_s\tau^+\tau^-$ may no longer be negligible and thus also have to be
examined. 

On the other hand, among rare B-meson decays, $B_s\to\tau^+\tau^-\gamma$ 
is of special interest due to its relative cleanliness and sensitivity 
to models beyond the SM \cite{Aliev97,Iltan00}.  We emphasize that when 
photon is emitted in additional to the lepton pair, no helicity suppression 
exists, and ``large''  branching ratio is expected.  As in the decay of 
$B_s\to\tau^+\tau^-$, we could expect that for $B_s\to\tau^+\tau^-\gamma$ 
the contributions from exchanging scalars could be sizable.

In this work, we will study the inclusive and exclusive decays 
$B\to X_s\tau^+\tau^-$ and $B_s\to\tau^+\tau^-\gamma$
in technicolor model with scalars. Since several scalars are predicted
in this model, they are expected to cause sizable effects in these
decays. Taking into account the contributions from both the neutral and 
charged scalars predicted in this model,  we will evaluate the branching 
ratios, the forward-backward asymmetry as well as the distributions of 
differential branching ratios.  This paper is organized as follows. 
Section \ref{sec:model} is a brief review of the model. The detailed 
calculations of the contributions from the scalars are presented in 
Section \ref{sec:bstt} and \ref{sec:bttg} for the decays 
$B\to X_s\tau^+\tau^-$ and $B_s\to\tau^+\tau^-\gamma$, respectively.
Finally, in  Sec.\ \ref{sec:result} we give some numerical results 
and conclusions. 
 
 \section{Technicolor with scalars}
\label{sec:model}

In the technicolor model with scalars, the rare decay processes we will
study receive contributions not only from the SM particles but also from
charged and neutral physical scalars predicted in such a  technicolor model. 
In this section we will briefly discuss the model and give the relevant 
Lagrangian which are needed in our calculation. More details of the model 
have been described in Refs.~\cite{Simmons89,Carone94}.

The gauge structure of the technicolor model with scalars is 
simply the direct product of the technicolor and standard model gauge
groups: ${SU(N)_{TC}\times SU(3)_{C}\times SU(2)_{W}\times U(1)_{Y}}$ 
\cite{Carone94}. The ordinary techni-singlet fermions are exactly as  
those in the SM. The technicolor sector consists of two 
techniflavors {\em p} and {\em m} that also transform under ${\em SU(2)_{W}}$.
In addition to the above particle spectrum, there exists a scalar doublet 
$\phi$ to which both the ordinary fermions and technifermions are coupled.
Unlike the SM Higgs doublet, $\phi$ does not cause electroweak symmetry 
breaking but obtains a non-zero effective vacuum expectation value (VEV)
when technicolor breaks the symmetry.

Writing the matrix form of the scalar doublet as
\begin{equation}
\Phi=\left[ \begin{array}{cc}
            \bar{\phi}^0 & \phi^+\\
            -\phi^-      & \phi^0
            \end{array} \right ]
\equiv \frac{(\sigma+f^{'})}{\sqrt{2}}\Sigma^{'},
\end{equation}
we then have the kinetic terms for the scalar fields given by
\begin{equation}
{\cal L}_{K.E.}=\frac{1}{2}\partial_\mu\sigma\partial^\mu\sigma+
\frac{1}{4}f^2Tr({D}_\mu\Sigma^\dagger {D}^\mu\Sigma)+
\frac{1}{4}(\sigma+f^{'})^2Tr({D}_\mu\Sigma^{'\dagger}{D}^\mu\Sigma^{'}).
\label{kinetic}
\end{equation}
Here the non-linear representation $\Sigma=exp(\frac{2i\pi}{f})$
and $\Sigma^{'}=exp(\frac{2i\pi^{'}}{f^{'}})$ are adopted for the technipion 
fields. $\sigma$ is an  isosinglet scalar field, $f$ and $f{'}$ are the 
technipion decay constant and the effective VEV, respectively.
The covariant derivative is defined as 
${D}^\mu\Sigma=\partial^\mu\Sigma-igW_a^\mu\frac{\tau^a}{2}\Sigma
+ig^{'} B^\mu\Sigma\frac{\tau^3}{2}$ with $\tau^a/2$ ($a=1,2,3$) being
the SU(2) generators, and  $W_\mu^a\ (B_\mu)$ denote the SU(2) (U(1)) vector 
fields with the gauge coupling constant $g$ ($g^{'}$). 
The definition of ${\em D}^\mu\Sigma^{'}$ is analogous to that of 
${\em D}^\mu\Sigma$.

The mixing between $\pi$ and $\pi^{'}$ gives 
\begin{eqnarray} 
\label{pia}
\pi_a&=&\frac{f\pi+f^{'}\pi^{'}}{\sqrt{f^2+f^{'2}}}, \\
\label{pip}
\pi_p&=&\frac{-f^{'}\pi+f\pi^{'}}{\sqrt{f^2+f^{'2}}},
\end{eqnarray}
with $\pi_a$ becoming the longitudinal component of the W and Z, 
and  $\pi_p$ remaining in the low-energy theory as an isotriplet of physical 
scalars. 
From Eq.\ (\ref{kinetic}) one can obtain the correct gauge boson masses
providing that $f^2+f^{'2}=V^2$ with the electroweak scale $V=246\ GeV$.

Additionally, the contributions to scalar potential generated by 
the technicolor interactions should be included in this model. The
simplest term one can construct is
\begin{equation}
{\cal L}_T=c_14\pi f^3Tr\left[\Phi\left(
\begin{array}{cc}
h_+ & 0\\
0 & h_-
\end{array}
\right)
\Sigma^\dagger\right] +h.c.,
\label{poential}
\end{equation}
where $c_1$ is a coefficient of order unity, $h_+$ and $h_-$ are the 
Yukawa couplings of scalars to $p$ and $m$ . 
From Eq.\ (\ref{poential}) the mass of the 
charged scalar at lowest order is obtained as  
\begin{equation}
m_{\pi_p}^2=2\sqrt{2}(4\pi f/f'){v}^2h
\end{equation}
with $h=(h_++h_-)/2$.  

In general, $f$ and $f'$ depend on $h_+,\ h_-,\ M_\phi$ and $\lambda$, 
where $M_\phi$ is the mass of the scalar doublet $\phi$, and  $\lambda$
is $\phi^4$ coupling. Two limits of the model have been studied previously
in the literatures: $(i)$ the limit in which $\lambda$ is small and can be
neglected \cite{Simmons89}, and $(ii)$ the limit in which $M_\phi$ 
is small and can be neglected \cite{Carone94}. When the largest 
Coleman-Weinberg corrections for the $\sigma$ field are included in 
the effective chiral Lagrangian \cite{Carone94}, $M_\phi,\ \lambda$ are 
replaced by the shifted scalar mass ${\tilde M}_\phi$ and coupling 
$\tilde\lambda$. In this case, one obtains the constraint
\begin{equation}
{\tilde M}_\phi^2 f'+\frac{\tilde \lambda}{2}f^{'3}=8\sqrt{2}\pi c_1hf^3
\end{equation}
and the isoscalar mass as
\begin{equation} 
m_\sigma^2={\tilde M}_\phi^2+\frac{2}{3\pi^2}[6(\frac{m_t}{f'})^4
+Nh^4]f^{'2}
\label{sigma1}
\end{equation}
in limit $(i)$, and 
\begin{equation} 
m_\sigma^2=\frac{3}{2}{\tilde \lambda}f^{'2}-\frac{1}{4\pi^2}
[6(\frac{m_t}{f'})^4+Nh^4]f^{'2}
\label{sigma2}
\end{equation}
in limit $(ii)$. 
The advantage of working in these two
limits is that at the lowest order the phenomenology depends on $h$, not on
the difference of $h_+$ and $h_-$, and can be described in terms 
of  (${\tilde M}_\phi, h)$ in limit $(i)$ and (${\tilde\lambda}, h)$ 
in limit $(ii)$. In this paper, we will work in the unitary gauge, where 
the particle spectrum consists of $\pi_p,\ \sigma$ and the massive 
weak gauge bosons. We choose two parameters $(f/f^{'},\ m_{\pi_p})$ in both 
limits of the model, and assume  $N=4,\ c_1=1$ in numerical 
calculations.

The interactions relevant to our calculations can be extracted from 
Eq.\ (\ref{kinetic}) and Eq.\ (\ref{poential}) and are given by
\begin{eqnarray}
{\cal L}&=&(\frac{f}{f^{'}})\frac{gm_W}{2}\sigma W_\mu^+W_\mu^-
+(\frac{f^{'}}{V})\frac{gm_Z}{\cos\theta_W}\sigma Z^\mu Z_\mu
-(\frac{f}{V})\frac{g}{2}\sigma\left[ W_\mu^-\partial^\mu\pi_p^+ + 
W_\mu^+\partial^\mu\pi_p^-\right ]\nonumber\\
&&+(\frac{V}{f^{'}})\frac{gm_{\pi_p}^2}{2m_W}\sigma\pi_p^+\pi_p^- 
+(\frac{V}{f^{'}})\frac{ig}{2m_W}\sigma
\left[m_U\overline{U} U+m_D\overline{D}D\right]\nonumber\\
&&+\frac{ig}{2}\left[W_\mu^-\pi_p^{+}\stackrel{\leftrightarrow}
{\partial^\mu}\pi_p^0+ W_\mu^+\pi_p^{-}\stackrel{\leftrightarrow}
{\partial^\mu}\pi_p^0\right]
+\frac{ig\cos 2\theta_W}{2\cos\theta_W}Z_\mu\pi_p^{+}
\stackrel{\leftrightarrow}{\partial^\mu}\pi_p^-\nonumber\\
&&-eA_\mu\pi_p^{+}\stackrel{\leftrightarrow}
{\partial^\mu}\pi_p^-
+(\frac{f}{f^{'}})\frac{ig}{2m_W}\pi_p^0\left[m_U\overline{U}\gamma_5 U
-m_D\overline{D}\gamma_5 D\right]\nonumber\\
&&-(\frac{f}{f^{'}})\frac{ig}{2\sqrt{2}m_W}
\left\{\pi_p^+\overline{U}_i[(m_U-m_D)-
(m_U+m_D)\gamma_5]V_{ij}D_j\right.\nonumber\\
&&\left.\ \ \ \ \ \ \ \ \ \ \ \  \ -\pi_p^-\overline{D}_iV_{ij}^*[(m_U-m_D)
+(m_U+m_D)\gamma_5]U_j\right\} \ ,
\end{eqnarray}
where $U,\ D$ and $m_U,\ m_D$ represent the column vector and the diagonal 
mass matrix for up and down-quarks, respectively. $\pi_p$ stands for the 
scalar field and ${\em V}_{ij}$ are the elements of 
the CKM matrix. The physical scalar-lepton couplings can be read off from the 
expression above by replacing $(U,D)$ with the corresponding lepton fields, 
replacing quark mass matrices with the corresponding diagonal lepton mass 
matrices, and setting $V_{ij}=1$.      

\section{$B\to X_s\tau^+\tau^-$ in technicolor model with scalars}
\label{sec:bstt}

It is well known that inclusive decay rates of heavy
hadrons can be calculated in  heavy quark effective theory
\cite{Neubert94}, and the leading terms in $1/m_Q$ expansion turn out to be
the decay of a free heavy quark and corrections stem from the order
$1/m_Q^2$\cite{Falk94}. In the technicolor model with scalars, the short 
distance contribution to $b\to s\tau^+\tau^-$ decay can 
be computed in the framework of the QCD corrected effective weak Hamiltonian,
obtained by integrating out heavy particles, i.e., top quark, scalar 
$\sigma,\ \pi_p$ and $W^\pm,\ Z$ bosons
\begin{equation}
{\cal H}_{eff}=\frac{4G_F}{\sqrt{2}}V_{tb}V_{ts}^*
\left(\sum_{i=1}^{10}[C_i(\mu){\cal O}_i(\mu)+C_{Q_i}(\mu){\cal Q}_i(\mu)]
\right) \ ,
\label{hamilton}
\end{equation}
where  ${\cal O}_i$ are the same as these given in Ref.\cite{Grinstein89}.
The additional operators ${\cal Q}_i$\cite{Dai97} are due to the neutral 
scalars exchange diagrams, which give considerable 
contributions in the case that the final lepton pair is $\tau^+\tau^-$.
Here we only present the explicit expressions of the operators 
governing $B\to X_s\tau^+\tau^-$. They read
\begin{eqnarray}
{\cal O}_7&=&\frac{e}{16\pi^2}m_b(\bar{s}_{\alpha}
\sigma^{\mu\nu}Rb_{\alpha})F_{\mu\nu},\nonumber\\
{\cal O}_8&=&\frac{e}{16\pi^2}(\bar{s}_{\alpha}
\gamma^{\mu}Lb_{\alpha})(\bar{\tau}\gamma_\mu\tau),\nonumber\\
{\cal O}_9&=&\frac{e}{16\pi^2}(\bar{s}_{\alpha}
\gamma^{\mu}Lb_{\alpha})(\bar{\tau}\gamma_\mu\gamma_5\tau),\nonumber\\
{\cal Q}_1&=&\frac{e^2}{16\pi^2}(\bar{s}_{\alpha}R
b_{\alpha})(\bar{\tau}\gamma_\mu\tau),\nonumber\\
{\cal Q}_2&=&\frac{e^2}{16\pi^2}(\bar{s}_{\alpha}R
b_{\alpha})(\bar{\tau}\gamma_\mu\gamma_5\tau) \ ,
\end{eqnarray}
where $L,R=(1 \mp \gamma_5)/2$,  $\alpha$ is the SU(3) color index
and  $F^{\mu\nu}$ the field strength tensor of the electromagnetic 
interaction.

In general in theories beyond the SM there will be additional contributions, 
which {\em are characterized by the values of the coefficients $C_{i}$
and $C_{Q_j}$ at the perturbative scale $m_W$}.  Using the Feynman rules 
presented in the preceding section, we can calculate the additional 
contributions arising from the scalars $\sigma$,  $\pi_p^0$ and 
$\pi_p^{\pm}$.  At the scale of $m_W$, the Feynman diagrams
for the charged scalar contributions are depicted in Fig.~1, while
Fig.~2 shows the additional contributions from the neutral scalars. 

The contributions of  Fig.~1  to the Wilson coefficients at  leading order 
read\footnote{ The contributions of $\pi_p^{\pm}$ take a 
similar form from those contributions of the color-singlet charged 
pseudo-Goldstone boson in the one-generation technicolor model \cite{Lu97}. 
The typical difference is the factor $f/f'$ in these new contributions.}   
\begin{eqnarray}
{C}_7(m_{W})_{TC}&=&(\frac{f}{f^{'}})^2{H}_1(x_{\pi_p}),\nonumber\\
{C}_8(m_{W})_{TC}&=&
(\frac{f}{f^{'}})^2\frac{4\sin^2\theta_W-1}{\sin^2\theta_W}
[x_W{H}_2(x_{\pi_p})-x_{\pi_p}H_3(x_{\pi_p})],\nonumber\\
{C}_9(m_{W})_{TC}&=&(\frac{f}{f^{'}})^2
\frac{x_W}{\sin^2\theta_W}H_2(x_{\pi_p}) \ ,
\label{CW} 
\end{eqnarray}
where $x_i=m_t^2/m_i^2$, $\theta_W$ is the Weinberg angle and the
functions $H_i$ can be expressed as 
\begin{eqnarray} 
{H}_1(x)&=&\frac{x}{12(x-1)^3}\left[
\frac{22x^2-53x+25}{6}-\frac{3x^2-8x+4}{x-1}\ln\ x\right],\nonumber\\
{H}_2(x)&=&\frac{x}{8(x-1)}\left[-1+
\frac{1}{x-1}\ln\ x\right],\nonumber \\
{H}_3(x)&=&\frac{1}{18(x-1)^3}\left[
\frac{47x^2-79x+38}{6}-\frac{3x^3-6x+4}{x-1}\ln\ x\right].
\end{eqnarray}

Theoretical calculations show that the contributions of Fig.~2 are significant
only for large $f/f^{'}$.  Keeping only the leading terms in large $f/f^{'}$ 
limit, the $C_{Q_j}(m_W)$ induced from these diagrams are given by 
\begin{eqnarray} 
C_{Q_1}(m_W)&=&-(\frac{f}{f^{'}})^4
\frac{m_bm_\tau}{m_\sigma^2}\frac{x_W}{4\sin^2\theta_W}H_4(x_{\pi_p}),
\nonumber\\
C_{Q_2}(m_W)&=&-(\frac{f}{f^{'}})^4\frac{m_bm_\tau}{m_{\pi_p}^2}
\frac{x_W}{4\sin^2\theta_W}H_5(x_{\pi_p}),
\end{eqnarray}
with
\begin{eqnarray} 
H_4(x)&=&\frac{1}{2(x-1)^2}\left[\frac{4x^2-7x+1}{2}
-\frac{x^2-2x}{x-1}\ln\ x\right],\nonumber\\
H_5(x)&=&\frac{1}{(x-1)}\left[\frac{x+1}{2}
-\frac{x}{x-1}\ln\ x\right].
\end{eqnarray}

It is noticeable that the contributions of Fig. 2 are proportional to 
$(f/f^{'})^4$, while those of  Fig. 1 proportional to 
$(f/f^{'})^2$. So for a sufficiently large $f/f^{'}$,
the contributions of neutral scalars in Fig. 2 are relatively enhanced
and become comparable with those from charged scalars in  Fig. 1.  

Neglecting the strange quark mass, the effective Hamiltonian 
(\ref{hamilton}) leads to the following matrix element for 
the inclusive $b\to s\tau^+\tau^-$ decay,
\begin{eqnarray}
{\cal M}&=&\frac{\alpha_{em}G_F}{2\sqrt{2}\pi}V_{tb}V_{ts}^*
\left\{-2C_7^{eff}\frac{m_b}{p^2}\bar{s}i\sigma_{\mu\nu}p_\nu(1+\gamma_5)b
\right.\nonumber\\
&&\left.+C_8^{eff}\bar{s}\gamma_\mu(1-\gamma_5) b\bar{\tau}\gamma_\mu\tau+
C_9\bar{s}\gamma_\mu(1-\gamma_5) b\bar{\tau}\gamma_\mu\gamma_5\tau
\right.\nonumber\\
&&\left.+C_{Q_1}\bar{s}(1+\gamma_5)b\bar{\tau}\tau
+C_{Q_2}\bar{s}(1+\gamma_5)b\bar{\tau}\gamma_5\tau\right\}.
\label{matrix}
\end{eqnarray}

The Wilson coefficients $C_i,\ C_{Q_j}$ are to be evaluated from $m_W$ down
to the lower scale of about $m_b$ by using the renormalization group equation.
When evolving down to b quark scale, the operators ${\cal O}_{1,2}$
and ${\cal Q}_3$ can mix with ${\cal O}_i,\  (i=7,8)$; however,
they can be included in an ``effective'' ${\cal O}_{7,8}$ because of 
their same structures contributing to the  $b\to s\tau^+\tau^-$ 
matrix element. At leading order, the Wilson coefficients are 
\cite{Grinstein89,Dai97,Hewett96}   
\begin{eqnarray} \label{c7}
C_7^{eff}(m_b)&=&\eta^{-16/23}\left\{C_7(m_W)-
\left[\frac{58}{135}(\eta^{10/23}-1)+\frac{29}{189}(\eta^{28/23}-1)\right]
C_2(m_W)-0.012C_{Q_3}(m_W)\right\},\\
 \label{c8}
C_8^{eff}(m_b)&=&C_8(m_W)+\frac{4\pi}{\alpha_s(m_b)}
\left[\frac{4}{33}(\eta^{-11/23}-1)+\frac{8}{87}(1-\eta^{-29/23})\right]
\nonumber\\
&&+\left\{g(\frac{m_c^2}{m_b^2},\frac{p^2}{m_b^2})-\frac{3\pi}{\alpha_{em}^2}
\kappa\sum_{V_i=\Psi^{'},\Psi^{''},\cdots}\frac{m_{V_i}\Gamma
(V_i\to \tau^+\tau^-)}{m_{V_i}^2-p^2-im_{V_i}\Gamma_{V_i}}\right\}
[3C_1(m_b)+C_2(m_b)],\\
C_9(m_b)&=&C_9(m_W),\\
C_{Q_i}(m_b)&=&\eta^{-\gamma_Q/\beta_0}C_{Q_i}(m_W).
\end{eqnarray}
Here $p$ is the momentum transfer, and 
\begin{equation} 
C_{Q_3}(m_W)=\frac{m_be^2}{m_\tau g^2}[C_{Q_1}(m_W)+C_{Q_2}(m_W)] \ ,
\end{equation}
where $\gamma_Q=-4$ is the anomalous dimension of $\bar{s}Rb,\ \beta_0=11-2n_f/3$, $\eta=\alpha_s(m_b)/\alpha_s(m_W)$, $C_2(m_W)=1$ and  
$C_{1,2}(m_b)=(\eta^{-6/23}\mp\eta^{12/23})/2$.   
$g(m_c^2/m_b^2,s)$ in  Eq.\ (\ref{c8}) arises from 
the one-loop matrix elements of the four-quark operators, and
\begin{equation}
g(x,y)=-\frac{4}{9}\ln\ x +\frac{8}{27}+\frac{16x}{9y}-
\frac{4}{9}(1+\frac{2x}{y}){\vert 1-
\frac{4x}{y}\vert}^{1/2}
\left\{
\begin{array}{cc}
\ln\ Z(x,y)-i\pi & for~~4x/y< 1\nonumber\\
2\arctan \frac{1}{\sqrt{4x/y-1}},&for~~4x/y> 1
\end{array}
\right.                                      
\end{equation}
where
\begin{equation}
Z(x,y)=\frac{1+\sqrt{1-\frac{4x}{y}}}{1-\sqrt{1-\frac{4x}{y}}} .
\label{zxy}
\end{equation}
The second term in brace of Eq.\ (\ref{c8}) estimates the 
long-distance contribution from the intermediate $\Psi^{'},\Psi^{''},\cdots$
\cite{Grinstein89}.  The phenomenological parameter $\kappa$ is taken as 
2.3\cite{Ali91} in our numerical calculations.

The formula of invariant dilepton mass distribution has been derived
in \cite{Dai97}, which is given by
\begin{eqnarray}\label{dgbtt}
\frac{d\Gamma(B\to X_s\tau^+\tau^-)}{ds}
=Br(B\to X_c\ell\nu)\frac{\alpha_{em}^2}{4\pi^2}
\vert\frac{V_{tb}V_{ts}^*}{V_{cb}}\vert^2f^{-1}(m_c/m_b)
(1-s)^2(1-\frac{4r}{s})^{1/2}D(s)
\end{eqnarray}
with 
\begin{eqnarray}
D(s)&=&4|C_7^{eff}|^2(1+\frac{2r}{s})(1+\frac{2}{s})+
|C_8^{eff}|^2(1+\frac{2r}{s})(1+2s)+|C_9|^2
(1-8r+2s+\frac{2r}{s})\nonumber\\
&&+12Re(C_7^{eff}C_8^{eff*})(1+\frac{2r}{s})
+\frac{3}{2}|C_{Q_1}|^2(s-4r)+\frac{3}{2}|C_{Q_2}|^2s
+6Re(C_9C_{Q_2}^*)r^{1/2}.
\end{eqnarray}
Here $s=p^2/m_b^2$, $r=m_\tau^2/m_b^2$. Function $f(x)=1-8x^2+8x^6-x^8
-24x^4\ln x$ is the phase-space factor. 

The angular information and the forward-backward  asymmetry are also 
sensitive to the details of the new physics. Defining the forward-backward  
asymmetry as
\begin{equation}
A_{FB}(s)=\frac{\int_0^1 d\cos\theta(d^2\Gamma/dsd\cos\theta)
-\int_{-1}^0 d\cos\theta(d^2\Gamma/dsd\cos\theta)}
{\int_0^1 d\cos\theta (d^2\Gamma/dsd\cos\theta)+\int_{-1}^0 d\cos\theta
(d^2\Gamma/dsd\cos\theta)} \ , 
\label{af}
\end{equation}    
where $\theta$ is the angle between the momentum of B-meson and 
$\tau^+$ in the center of mass frame of the dilepton, we obtain
\begin{equation}
A_{FB}=\frac{6(1-4r/s)^{1/2}}{D(s)}Re(2C_7^{eff}C_9^*+C_8^{eff}C_9^*s
+ 2C_7^{eff}C_{Q_2}^*r^{1/2}+C_8^{eff}C_{Q_1}^*r^{1/2}).
\end{equation}

\section{$B_s\to \tau^+\tau^-\gamma$ in technicolor model with scalars}
\label{sec:bttg}

Now let us turn to rare radiative decay  
$B_s\to \tau^+\tau^-\gamma$. The  exclusive decay can be obtained 
from the inclusive decay $b\to s\tau^+\tau^-\gamma$, and further, 
from $b\to s\tau^+\tau^-$. To achieve this, it is necessary to attach
photon to any charged internal and external lines in the Feynman diagrams
of $b\to s\tau^+\tau^-$. As pointed out in Ref.\ \cite{Aliev97},
contributions coming from the attachment of photon to any charged internal 
line are strongly suppressed and we can neglect them safely. However,
since the mass of $\tau$-lepton is not much smaller than that of $B_s$-meson, 
in $B_s\to \tau^+\tau^-\gamma$ decay, the contributions of the 
diagrams with photon radiating from  final leptons are  comparable  with  
those from initial quarks.  When a photon is attached to the initial quark 
lines, the corresponding matrix element for the $B_s\to
\tau^+\tau^-\gamma$ decay can be written as 
\begin{eqnarray}
{\cal M}_1=\frac{\alpha_{em}^{3/2}G_F}{\sqrt{2\pi}}V_{tb}V_{ts}^*
&&\left\{[A\varepsilon_{\mu\alpha\beta\sigma}\epsilon_\alpha^*p_\beta q_\sigma
+iB(\epsilon_\mu^*(pq)-(\epsilon^*p)q_\mu)]\bar{\tau}\gamma_\mu\tau\right.
\nonumber\\
&&+\left.[C\varepsilon_{\mu\alpha\beta\sigma}\epsilon_\alpha^*p_\beta q_\sigma
+iD(\epsilon_\mu^*(pq)-(\epsilon^*p)q_\mu)]\bar{\tau}\gamma_\mu\gamma_5\tau
\right\},
\label{M1}
\end{eqnarray}
where 
\begin{eqnarray}
A&=&\frac{1}{m_{B_s}^2}[C_8^{eff}g_1(p^2)-2C_7\frac{m_b}{p^2}g_2(p^2)],\nonumber\\
B&=&\frac{1}{m_{B_s}^2}[C_8^{eff}f_1(p^2)-2C_7\frac{m_b}{p^2}f_2(p^2)],\nonumber\\
C&=&\frac{C_9}{m_{B_s}^2}g_1(p^2),\nonumber \\
D&=&\frac{C_9}{m_{B_s}^2}f_1(p^2).
\end{eqnarray}
In obtaining Eq.\ (\ref{M1}) we have used  
\begin{eqnarray}
\langle\gamma|\bar{s}\gamma_\mu(1\pm\gamma_5)|B_s\rangle
=&\frac{e}{m_{B_s}^2}\left\{\varepsilon_{\mu\alpha\beta\sigma}\epsilon_\alpha^*
p_\beta q_\sigma g_1(p^2)\mp i[(\epsilon_\mu^*(pq)-(\epsilon^*p)q_\mu)]f_1(p^2)
\right\},\\
\label{gmb1} 
\langle\gamma|\bar{s}i\sigma_{\mu\nu}p_\nu(1\pm\gamma_5)b|B_s\rangle
=&\frac{e}{m_{B_s}^2}\left\{\varepsilon_{\mu\alpha\beta\sigma}\epsilon_\alpha^*
p_\beta q_\sigma g_2(p^2)\pm i[(\epsilon_\mu^*(pq)-(\epsilon^*p)q_\mu)]f_2(p^2)
\right\},
\label{gmb2} 
\end{eqnarray}
and 
\begin{equation}
\langle\gamma|\bar{s}(1\pm\gamma_5)|B_s\rangle=0.
\label{gmb3} 
\end{equation}
Here $\epsilon_\mu$ and $q_\mu$ are the four vector polarization and momentum 
of photon, respectively; $g_i,\ f_i$ are form factors
\cite{Buchalla93,Eilam95}. 
Eq.\ (\ref{gmb3}) can be obtained by multiplying\ $p_\mu$ in both sides of 
Eq.\ (\ref{gmb1}) and using the equations of motion.  From Eq.\ (\ref{gmb3}) 
one can see that the neutral scalars do not contribute to the matrix element 
${\cal M}_1$.

When a photon is radiated from the final $\tau$-leptons, the situation is 
different. Using the expressions
\begin{eqnarray}
& & \langle 0|\bar{s}b|B_s\rangle=0,\nonumber\\
& & \langle 0|\bar{s}\sigma_{\mu\nu}(1+\gamma_5)b|B_s\rangle =0,\nonumber\\
& & \langle 0|\bar{s}\gamma_\mu\gamma_5|B_s\rangle =-if_{B_s}P_{B_s\mu}
\end{eqnarray}
and the conservation of the vector current, one finds that only  
the operators ${\cal Q}_{1,2}$ and ${\cal O}_9$ give contribution to this 
Bremsstrahlung part. The corresponding matrix is given by \cite{Iltan00}
\begin{eqnarray}
{\cal M}_2&=&\frac{\alpha_{em}^{3/2}G_F}{\sqrt{2\pi}}V_{tb}V_{ts}^*
i2m_\tau f_{B_s}\left\{(C_9+\frac{m_{B_s}^2}{2m_\tau m_b}C_{Q_2})\bar{\tau}
\left[\frac{\not\epsilon\not P_{B_s}}{2p_1q}-
\frac{\not P_{B_s}\not\epsilon}{2p_2q}\right]\gamma_5\tau\right.\nonumber\\
&&+\left.\frac{m_{B_s}^2}{2m_\tau m_b}C_{Q_1}
\left[2m_\tau(\frac{1}{2p_1q}+\frac{1}{2p_2q})\bar{\tau}\not\epsilon\tau+
\bar{\tau}(\frac{\not\epsilon \not P_{B_s}}
{2p_1q}-\frac{\not P_{B_s}\not\epsilon}{2p_2q})\gamma_5\tau\right]\right\}.
\end{eqnarray}
Here  $P_{B_s},\ f_{B_s}$ are the momentum and the decay constant of 
the $B_s$ meson. 

Finally, the total matrix element for the $B_s\to \tau^+\tau^-\gamma$
decay is obtained as a sum of the ${\cal M}_1$ and ${\cal M}_2$. After 
summing over the spins of the $\tau$-leptons and polarization of the photon, 
we get the square of the matrix element as
\begin{equation}
\vert{\cal M}\vert^2=\vert{\cal M}_1\vert^2+\vert{\cal M}_2\vert^2+
2Re({\cal M}_1{\cal M}_2^*)
\end{equation}
with\footnote {There are some errors in Eqs.~(15) and (17) of Ref.\cite{Iltan00}. 
We believe the expressions of $C_{Q_1}^H(m_W)$ and $C_{Q_2}^H(m_W)$ are not 
correct in Two-Higgs-Double model I.}     
\begin{eqnarray}
\vert{\cal M}_1\vert^2&=&
4\vert\frac{\alpha_{em}^{3/2}G_F}{\sqrt{2\pi}}V_{tb}V_{ts}^*\vert^2
\left\{[\vert A\vert^2+\vert B\vert^2\vert][p^2((p_1q)^2+(p_2q)^2)
+2m_\tau^2(pq)^2]\right.\nonumber\\
&&+\left.[\vert C\vert^2+\vert D\vert^2\vert][p^2((p_1q)^2+(p_2q)^2)
-2m_\tau^2(pq)^2]\right.\nonumber\\
&&\left.+2Re(B^*C+A^*D)p^2((p_1q)^2-(p_2q)^2)\right\},
\end{eqnarray}
\begin{eqnarray}
2Re({\cal M}_1{\cal M}_2^*)&=&
-16\vert\frac{\alpha_{em}^{3/2}G_F}{\sqrt{2\pi}}V_{tb}V_{ts}^*\vert^2
m_\tau^2f_{B_s}(pq)^2\left\{
\vert C_9+\frac{m_{B_s}^2C_{Q_2}}{2m_\tau m_b}\vert\left[Re(A)
\frac{(p_1q+p_2q)}{(p_1q)(p_2q)}\right.\right.\nonumber\\
&&\left.\left.-Re(D)\frac{(p_1q-p_2q)}
{(p_1q)(p_2q)}\right]+ Re(B)\vert\frac{m_{B_s}^2C_{Q_1}}{2m_\tau m_b}\vert
\left[\frac{3m_{B_s}^2+2m_\tau^2-5(pq)}{(p_1q)(p_2q)}-
\frac{2p^2}{(pq)^2}\right]\right.\nonumber\\
&&\left.+Re(C)\vert\frac{m_{B_s}^2C_{Q_1}}{2m_\tau m_b}\vert
\left[\frac{(p_1q-p_2q)}{(p_1q)(p_2q)}(1+\frac{2p^2}{(pq)^2})\right]\right\},
\end{eqnarray}
\begin{eqnarray}
\vert{\cal M}_2\vert^2&=&
-8\vert\frac{\alpha_{em}^{3/2}G_F}
{\sqrt{2\pi}}V_{tb}V_{ts}^*\vert^2m_\tau^2f_{B_s}^2
\left\{\vert C_9+\frac{m_{B_s}^2C_{Q_2}}{2m_\tau m_b}\vert^2
\left[\frac{m_\tau^2m_{B_s}^2(pq^2)}{(p_1q)^2(p_2q)^2}-
\frac{m_{B_s}^2p^2+2(pq)^2}{(p_1q)(p_2q)}\right]
\right.\nonumber\\
&&\left.-\vert\frac{m_{B_s}^2C_{Q_1}}{2m_\tau m_b}\vert^2
\left[\frac{m_\tau^2(m_{B_s}^2-4m_\tau^2)(pq)^2}{(p_1q)^2(p_2q)^2}
-\frac{(m_{B_s}^2-4m_\tau^2)p^2+2(pq)^2}{(p_1q)(p_2q)}\right]
\right\}.
\end{eqnarray}
Here $p_1,\ p_2$ are momenta of the final $\tau$-leptons. It is obvious
that the quantity $\vert{\cal M}\vert^2$ depends only on the 
scalar products of the momenta of the external particles. 

In the rest frame of the $B_s$, the photon energy $E_\gamma$ and the lepton 
energy $E_1$ are  restricted by
\begin{eqnarray}
0&&\leq E_\gamma\leq\frac{m_{B_s}^2-4m_\tau^2}{2m_{B_s}},\nonumber\\
\frac{m_{B_s}-E_\gamma}{2}-\frac{E_\gamma}{2}
\sqrt{1-\frac{4m_\tau^2}{m_{B_s}^2-2m_{B_s}E_\gamma}}&&\leq E_1\leq
 \frac{m_{B_s}-E_\gamma}{2}+\frac{E_\gamma}{2}
\sqrt{1-\frac{4m_\tau^2}{m_{B_s}^2-2m_{B_s}E_\gamma}}.
\end{eqnarray}
However, in $\vert{\cal M}_2\vert^2$ it appears an infrared divergence, 
which originates in the Bremsstrahlung processes when photon
is soft and in this case, the $B_s\to\tau^+\tau^-\gamma$ 
can not be distinguished from $B_s\to\tau^+\tau^-$. 
Therefore, both processes must be considered together in order 
to cancel the infrared divergence. Taking the fact that the infrared 
singular terms in $\vert{\cal M}_2\vert^2$ exactly cancel the 
$O(\alpha_{em})$ virtual correction in $B_s\to\tau^+\tau^-$
amplitude in account\cite{Aliev97}, we follow Ref.\ \cite{Aliev97} and 
consider the photon in  $B_s\to\tau^+\tau^-\gamma$ 
as a hard photon and impose a cut on the photon energy $E_\gamma$, which 
correspond to the radiated photon can be detected in  the experiments. 
This cut requires  $E_\gamma\geq \delta~ m_{B_s}/2$ with $\delta = 0.02$.

After integrating over the phase space and the lepton 
energy $E_1$, we express the decay rate as
\begin{eqnarray}
\Gamma&=&\vert\frac{\alpha_{em}^{3/2}G_F}{2\sqrt{2\pi}}V_{tb}V_{ts}^*\vert^2
\frac{m_{B_s}^5}{(2\pi)^3}\left\{
\frac{m_{B_s}^2}{12}\int^{1-\delta}_{4\hat{r}}(1-\hat{s})^3d\hat{s}\sqrt{1-
\frac{4\hat{r}}{\hat{s}}} 
[(|A|^2+|B|^2)(\hat{s}+2\hat{r})\right.\nonumber\\
&&\left.+(|C|^2+|D|^2)(\hat{s}-4\hat{r})]
-2f_{B_s}\vert C_9+\frac{m_{B_s}^2C_{Q_2}}{2m_\tau m_b}\vert
\hat{r}\int^{1-\delta}_{4\hat{r}}(1-\hat{s})^2d\hat{s} Re(A)
\ln\ \hat{z}\right.\nonumber\\
&&-2f_{B_s}\vert\frac{m_{B_s}^2C_{Q_1}}{2m_\tau m_b}\vert
\hat{r}\int^{1-\delta}_{4\hat{r}}(1-\hat{s}) d\hat{s} Re(B)
\left[(1+4\hat{r}+5\hat{s})\ln\ \hat{z}+
\hat{s}\sqrt{1-\frac{4\hat{r}}{\hat{s}}}\right]\nonumber\\
&&\left.-\frac{4f_{B_s}^2}{m_{B_s}^2}
\vert C_9+\frac{m_{B_s}^2C_{Q_2}}{2m_\tau m_b}\vert^2\hat{r}
\int^{1-\delta}_{4\hat{r}}d\hat{s}\left[(1+\hat{s}+
\frac{4\hat{r}-2}{1-\hat{s}})\ln\ \hat{z}+
\frac{2\hat{s}}{1-\hat{s}}
\sqrt{1-\frac{4\hat{r}}{\hat{s}}}\right]\right.\nonumber\\
&&\left.+f_{B_s}^2
\vert\frac{C_{Q_1}}{m_b}\vert^2
\int^{1-\delta}_{4\hat{r}}d\hat{s}\left[(1-8\hat{r}+\hat{s}
-\frac{2-10\hat{r}+8\hat{r}^2}
{1-\hat{s}})\ln\ \hat{z}+\frac{2(1-4\hat{r})\hat{s}}{1-\hat{s}}
\sqrt{1-\frac{4\hat{r}}{\hat{s}}}\right]\right\} \ ,
\end{eqnarray}
where $\hat{s}=p^2/m_{B_s}^2,\ \hat{r}=m_\tau^2/m_{B_s}^2$. 
$\hat{z}\equiv Z(\hat{r},\hat{s})$ takes the form given in Eq.\ (\ref{zxy}).

\section{Numerical results and conclusion}
\label{sec:result}

In this section we give some numerical results and discussions. For reference, 
we present our SM predictions\footnote{Switching off the scalars contributions,
our formula for $Br(B_s\to \tau^+\tau^-)$ 
is the same as given in Ref.\ \cite{Aliev97}, but different from their result. 
In fact, the Bremsstrahlung part is $3.98\times 10^{-8}$ for $\delta=0.02$.}
\begin{eqnarray}
Br(B\to X_s\tau^+\tau^-)&=&2.59\times 10^{-6},\nonumber\\
Br(B_s\to \tau^+\tau^-\gamma)&=&5.18\times 10^{-8}.
\end{eqnarray}
These values are obtained for the fixed input parameters \cite{PDG00} 
listed in Table I and the QCD coupling constant 
$\alpha_s(m_b)=0.218$  which is calculated via
\begin{equation}
\alpha_s(\mu)=\frac{\alpha_s(m_Z)}
{1-\beta_0\frac{\alpha_s(m_Z)}{2\pi}\ln\frac{m_Z}{\mu}} 
\end{equation}  
with $\alpha_s(m_Z)=0.119$\cite{PDG00} and $m_Z=91.19\ GeV$.

%%%%%%%%%%%%%%%%%%%%%%%%
\null\vspace{0.4cm}
\noindent
{\small Table I. 
  The value of the input parameters used in the numerical calculations
 (mass and decay constant in unit GeV)}.
\begin{center}
\begin{tabular}{cccccc}  
\hline
$m_t$& $m_c$&$m_b$&$m_\tau$ & $m_{B_s}$& $m_W$\\
\hline
176  & 1.4 & 4.8 & 1.78 & 5.26  & 80.448 \\
\hline \hline								
$f_{B_s}$\cite{Belyaev95}&~~$|V_{tb}V_{ts}^*|$~~ & 
~~$|V_{tb}V_{ts}^*/V_{cb}|^2$~~
&$\alpha_{em}^{-1}$&$\tau(B_s)$&$\sin^2\theta_W$\\
\hline
0.14& 0.045&0.95&137&$~~1.64\times 10^{-12} s~~$&0.2325\\ \hline
\end{tabular}
\end {center}
\vspace{1cm}
%%%%%%%%%%%%%%%%%%%%%%%%

In addition, we use the masses, decay widths and branching rates of $J/\Psi$ 
family in Ref.\ \cite{PDG00}, the normalized factor, 
branching rate $Br(B\to X_c\ell\nu)=10.2\%$, and take the dipole 
forms of the form-factors given by Ref. \cite{Eilam95}
\begin{eqnarray}
g_1(p^2)&=&\frac{1\ GeV}{(1-p^2/5.6^2)^2},\  \ \ \  
g_2(p^2)=\frac{3.74\ GeV}{(1-p^2/40.5)^2},\nonumber\\
f_1(p^2)&=&\frac{0.8\ GeV}{(1-p^2/6.5^2)^2}, \ \ \ \
f_2(p^2)=\frac{0.68\ GeV}{(1-p^2/30)^2} . 
\label{formf}
\end{eqnarray}

As mentioned in Sec.\ \ref{sec:model}, in  limit (i) and (ii)  there are 
only two independent parameters  $f/f^{'}$ and $m_{\pi_p}$ in technicolor 
model with scalars. The limit on  $f/f^{'}$ can be obtained from
the studies of $b\to X_c\tau\nu$ \cite{Xiong99},  
$b\to X_s\gamma$ \cite{Xiong99,Carone95}, $Z\to b\bar{b}$ \cite{Carone95},
$B\to X_s\mu^+\mu^-$ \cite{Su97} and B-$\overline{B}$ mixing in technicolor 
model with scalars\cite{Simmons89,Carone94,Carone95}, which is given by
\cite{Xiong99}
\begin{equation}
\frac{f}{f^{'}}\leq 0.03 \left (\frac{m_{\pi_p}}{1\ GeV} \right ) 
                                     \ \ \  (95\%\ C.\ L.)
\label{ffpmpi}
\end{equation}  
This indicates that $f/f'$ could be large for large $m_{\pi_p}$. On the other 
hand, we should notice the induced values of $m_{\sigma}$ in 
Eqs.~(\ref{sigma1}, \ref{sigma2}). Although the lower experimental 
bound of $107.7\ GeV$ \cite{PDG00} on the SM Higgs boson may not be applied 
directly to  $m_{\sigma}$, we ensure that the neutral scalar $\sigma$ is not 
lighter than this value in our numerical calculations.

We found that the behaviors of the quantities we studied 
are similar in limit $(i)$ and $(ii)$. So for illustrations, we only present 
some numerical results in limit $(ii)$. Some numerical examples are presented 
in Figs. \ref{btt}-\ref{dbtt-s}.

Fig.~\ref{btt} and Fig.~\ref{bttg} show the branching ratios of 
$B\to X_s\tau^+\tau^-$ and $B_s\to\tau^+\tau^-\gamma$,
respectively. As expected, the branching ratios drop with $m_{\pi_p}$
and increase with $f/f'$. For the specified values of 
$m_{\pi_p}$ and $f/f'$ in the figures, the branching ratios  
can be enhanced over the SM results by a couple of orders of magnitude.
  
Fig.~\ref{af-s} show the forward-backward (FB) asymmetry 
as a function of the scaled invariant dilepton mass squared $s$.
We should point out that since  some common contributions appearing in both 
the numerator and the denominator cancel out to some extent, the FB asymmetry
is a sensitive, relatively model-independent probe of these models
\footnote{Although the FB asymmetry is less model-dependent for this reason, 
it is not completely model-independent since the model-dependent contributions 
cannot completely cancel out.}.    
Fig.~\ref{af-s} shows a significant difference 
between the SM and the technicolor predictions, especially in the 
region of large invariant dilepton mass.
One can see that, unlike the branching ratio, the forward-backward asymmetry 
is enhanced  significantly due to the neutral scalars contributions.

The differential branching ratios of  $B\to X_s\tau^+\tau^-$ and 
$B_s\to\tau^+\tau^-\gamma$ versus the scaled invariant dilepton mass 
squared are shown in Fig.~\ref{dbtt-s}. such distributions also 
differ significantly from the SM predictions. We stress that all these 
distributions would be useful for fitting the future experimental results 
in the framework of such a technicolor model, especially when some deviations 
from the SM predictions are discovered in future experiments.
Different models, such as technicolor models and 2HDM's, may all predict
same enhancements in branching ratios, but they may give different behaviors
for some distributions. To claim a given model is experimentally favored 
or disfavored, all these distributions would be useful.

We realized that since there are new model assumptions introduced 
here, such as the form-factor (\ref{formf}), as well as a large 
number of experimental inputs, each of which comes with its own 
uncertainty, our conclusions may only be qualitatively reliable. 
These are:

1.  The branching ratios  predicted by the technicolor model
    with scalars can be enhanced over the SM results by a couple of orders of 
    magnitude in parameter space we studied.

2.  The dominant contributions are from the exchange of charged scalars as 
    shown in Fig.\ 1.  This could be well understood since the contributions 
    from the charged scalars, in contrast to the decays studied in 
    2HDM-II\cite{Iltan00}, are not suppressed but enhanced by a factor 
    $(f/f^{'})^2$ when $f/f^{'}$ is large.

3.  The neutral scalars play no important role except for sufficiently 
    large  $f/f^{'}$. For the illustrative values in the numerical 
    results, i.e.,  $10\le f/f^{'}\le 15$, the contributions of    
    neutral scalars are still much smaller than those of charged scalars.
    However, as showed in the analytical expressions, the contributions of    
    neutral scalars in Fig.~2 are two orders higher in $f/f^{'}$  than 
    those of charged scalars in Fig.~1. Therefore, for sufficiently large 
    $f/f^{'}$ the contributions of neutral scalars could also be important.

4.   The possible large enhancements over those predicted by the SM 
     might be detectable in future experiments since the sensitivity of 
     the new generation of B experiment to these processes should be 
     quite high. If such large enhancements
     are observed in future experiments, they could be interpreted in such 
     a technicolor model although this model might not be the unique one to 
     explain them. If not observed, further stringent constraints on the model 
     parameter space could be obtained and thus this model would be severely
     disfavored.

\section*{Acknowledgment}

We thank C.-H. Chang and H. J. Yang for discussions and comments. 
This work is supported in part by a grant of Chinese Academy of Science
for Outstanding Young Scholars.

\newpage
\begin{center}

\begin{figure}
\epsfig{file=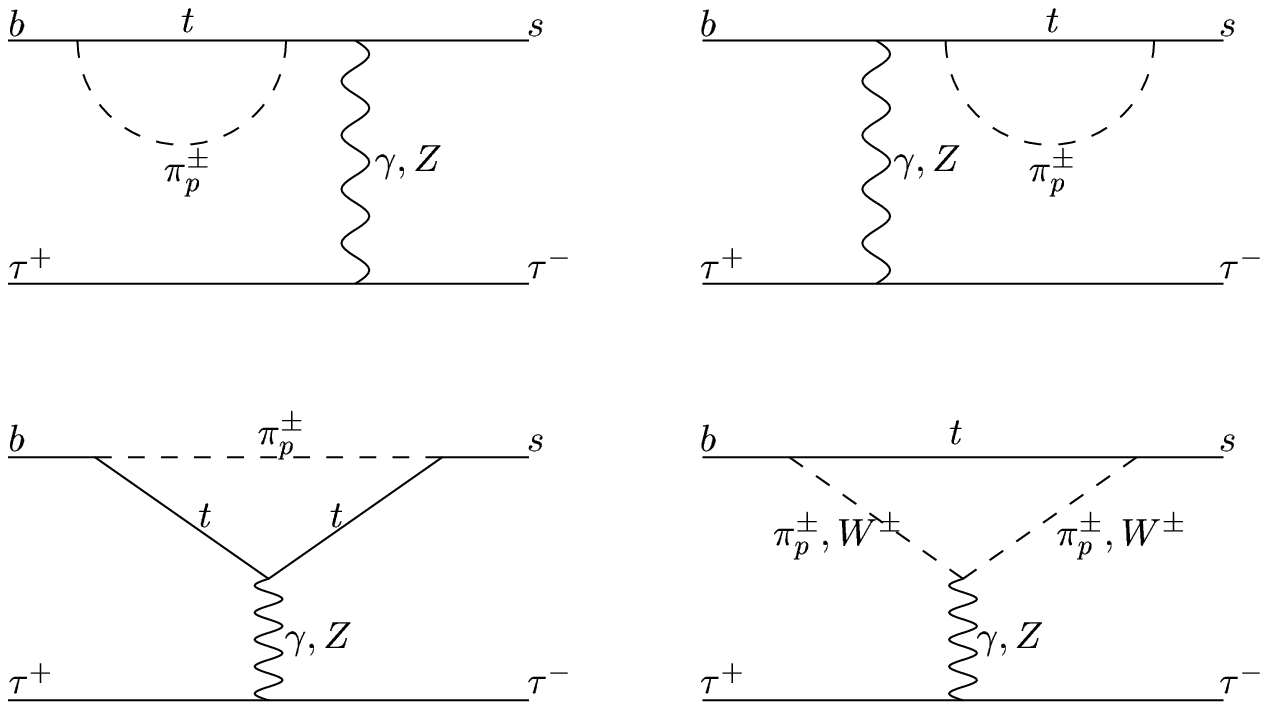 ,width=14cm}              
\caption{ Feynman diagrams for the charged scalar contributions 
           in technicolor with scalars.}
\end{figure}

\begin{figure}
\epsfig{file=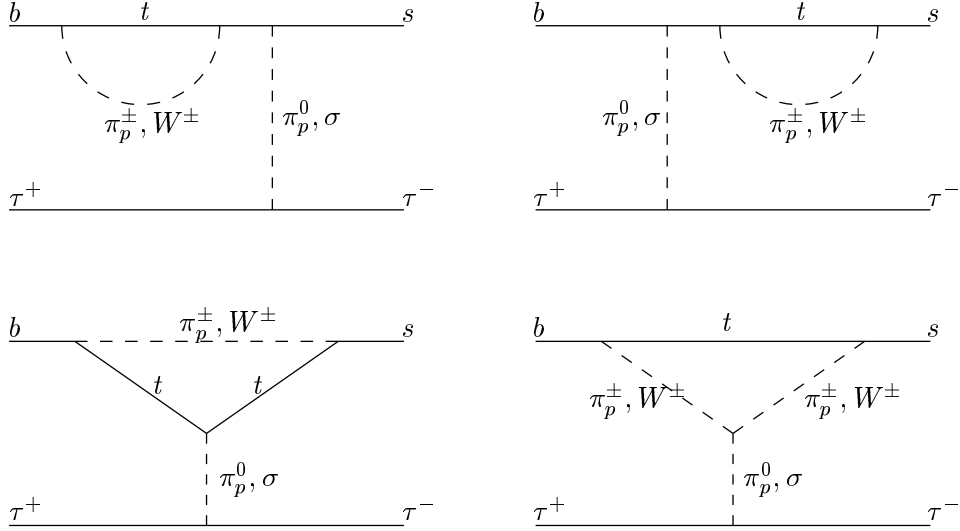 ,width=14cm}              
\caption{ Feynman diagrams for the additional contributions 
          from the neutral scalars in technicolor with scalars.}
\end{figure}
                                                          
\newpage
\begin{figure}
\epsfig{file=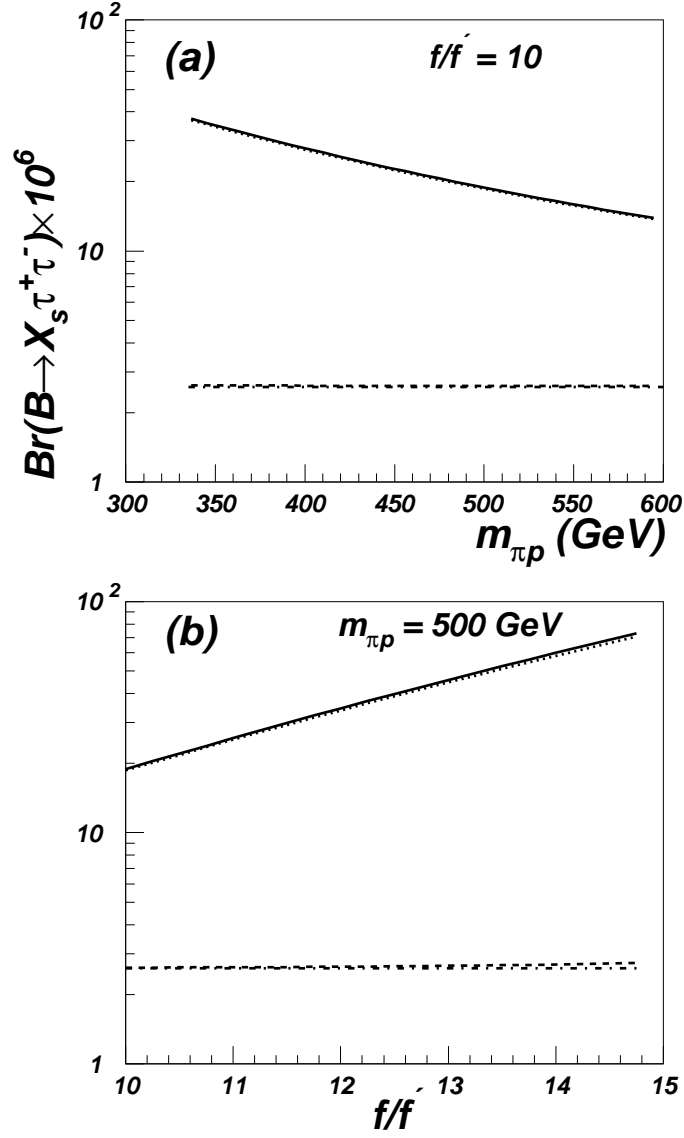 ,width=10cm}              
\caption{Branching ratio of $B\to X_s\tau^+\tau^-$ versus $m_{\pi_p}$ for 
$f/f^{'}=10$ (a),  and versus $f/f^{'}$ for $m_{\pi_p}=500~GeV$ (b).
The dot-dashed line stands for the SM prediction.  
The dotted (dashed) lines denote the new physics contributions from 
$\gamma,~Z$ exchange (neutral scalar exchange) diagrams shown in Fig.~1
(Fig.~2). The solid one is the total values.}
\label{btt}
\end{figure}
 
\begin{figure}
\epsfig{file=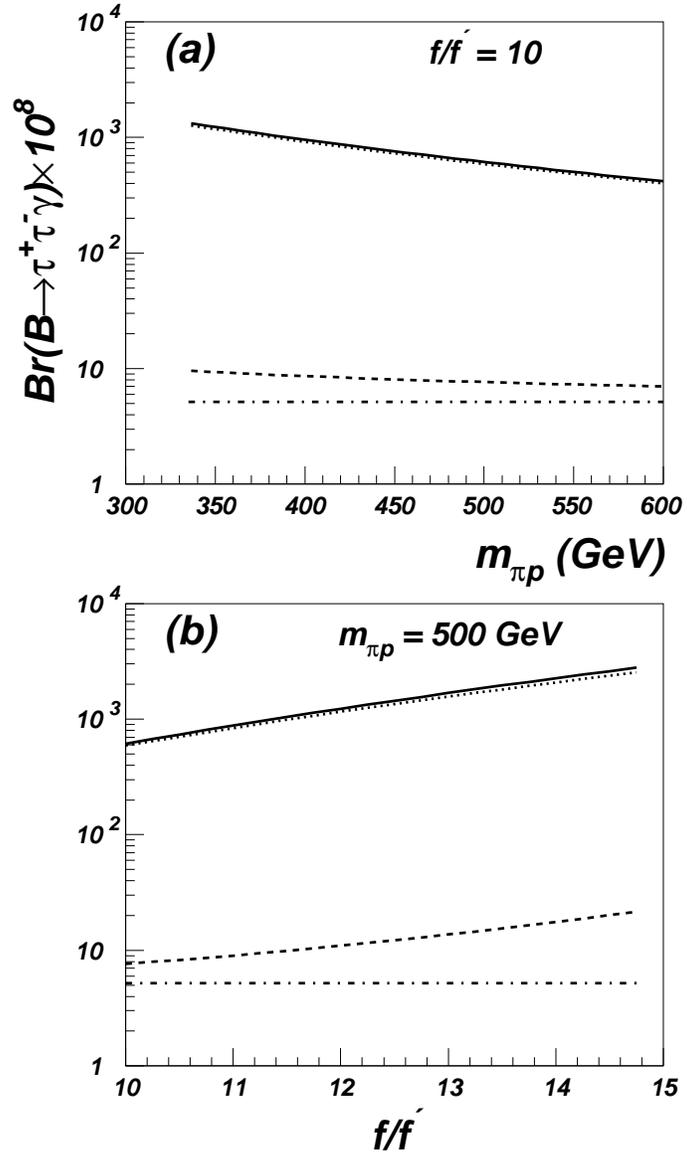 ,width=10cm}              
\caption{The same as Fig.~\ref{btt}, but for $B_s\to\tau^+\tau^-\gamma$.}
\label{bttg}
\end{figure}

\begin{figure}
\epsfig{file=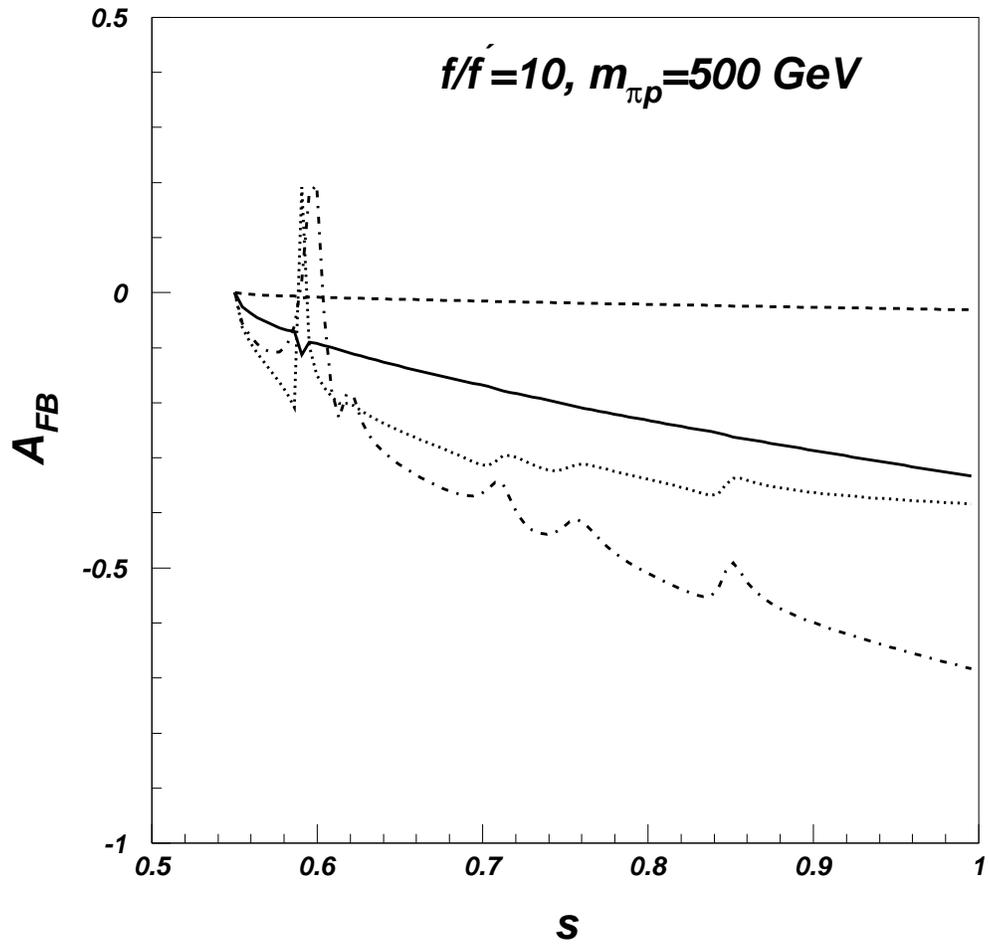 ,width=14cm}  
\caption{The same as Fig.~\ref{btt}, but for the forward-backward asymmetry
of $B\to X_s\tau^+\tau^-$ versus the scaled invariant dilepton mass squared
$s$ with $f/f^{'}=10$ and $m_{\pi_p}=500~GeV$.}
\label{af-s}
\end{figure}

\begin{figure}
\epsfig{file=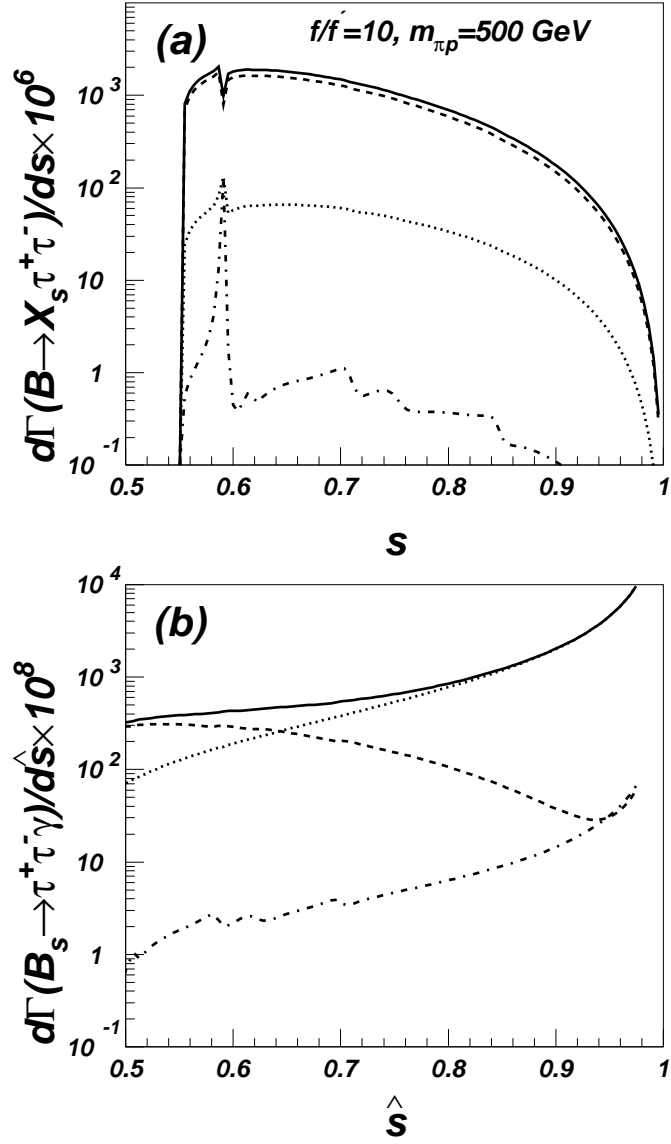 ,width=10cm} 
\caption{ The same as Fig.~\ref{btt}, but for the differential branching 
ratios of $B\to X_s\tau^+\tau^-$ and $B_s\to\tau^+\tau^-\gamma$
versus the scaled invariant dilepton mass squared  with $f/f^{'}=10$ and 
$m_{\pi_p}=500~GeV$. }
\label{dbtt-s}
\end{figure}

\end{center}

\end{document}